\documentclass[%
 reprint,
 amsmath,amssymb,
 aps,
 prl,
]{revtex4-1}
\usepackage{color}
\makeatletter
\usepackage{graphicx}
\usepackage{dcolumn}
\usepackage{bm}
\usepackage{float}
\usepackage{gensymb,comment}

\begin{document}


\title{Fingerprinting Heatwaves and Cold Spells and Assessing Their Response to Climate Change using Large Deviation Theory} 

\author{Vera Melinda Galfi}
\affiliation{Department of Earth Sciences, Uppsala University, Uppsala, Sweden}
\email{vera.melinda.galfi@geo.uu.se}
\author{Valerio Lucarini}%
\affiliation{Department of Mathematics and Statistics \& Centre for the Mathematics of Planet Earth, University of Reading, UK
}
\email{Corresponding author. Email: v.lucarini@reading.ac.uk}
%

\date{\today}

\begin{abstract}
Extreme events provide relevant insights {\color{black}into} the dynamics of  climate   and their understanding is key 
for mitigating the impact of climate variability and climate change. By applying large deviation theory to a state-of-the-art Earth system model, we define the climatology of persistent heatwaves and cold spells in key target geographical regions 
by {\color{black}estimating} 
the  rate functions for the surface temperature, and we assess the impact of increasing  CO$_2$ concentration on such persistent anomalies. Hence, 
we can better {\color{black}quantify} the increasing hazard {\color{black}due} to heatwaves in a warmer climate. We show that two 2010 high impact events - summer Russian heatwave and winter Dzud in Mongolia - are associated with atmospheric patterns that are exceptional compared to the typical ones, but typical compared to the climatology of extremes. {\color{black}Their dynamics is encoded in the natural variability of the climate.} Finally, we propose {\color{black}and test} an approximate formula for the return times of  large and persistent temperature fluctuations from easily accessible statistical properties. 
\end{abstract}

\maketitle


\paragraph{Introduction}Understanding extreme events is a key scientific challenge 
and is essential for addressing the natural hazards due to climate variability and climate change. High-impact events 
are usually associated 
with long temporal persistence, as resilience 
against anomalous environmental conditions does not last indefinitely \cite{Easterling2000,Who2004,Poumadere2005,Ipcc2012}. The theory of low-frequency variability of the atmosphere shows that long temporal persistence  and large spatial extent of patterns go hand in hand \cite{Ghil2002,Ghil2020}.
{\color{black}Let's consider two high-impact climatic extremes that occurred in 2010.} 
The summer Russian heatwave (RHW) 
had a temporal duration of about one month and a spatial extent of several million  Km$^2$ \cite{Barriopedro2011}. The winter Mongolian Dzud (MD)- an extreme cold spell - affected Mongolia and a large part of Siberia also for about a month  \cite{Rao2015,Sternberg2017}. Dzuds have historically been major drivers of migration for central Asian nomadic populations \cite{Fang92,Hvistendahl12}. Persistent large scale atmospheric patterns can have a cascade effect \cite{Kornhuber2019}: {\color{black}it is well known that }the 2010 RHW was dynamically linked \cite{Hong2011,Lau2012,Boschi2019} to extensive floods in Pakistan.

{\color{black}Future} changes in the statistics of heatwaves are  worrying, as more persistent positive temperature fluctuations compound with the trend in the average temperature \cite{Seneviratne2006,Coumou2013,Pfleiderer2019,Kornhuber2021}. Climate change leads to less frequent cold spells \cite{Smith2020}, even {\color{black}if specific dynamical} processes might 
facilitate their occurrence \cite{Kretschmer2018,Cohen2018}. {\color{black}Following \cite{Allen2003}, a lot of research has focused on understanding whether it is possible to attribute {\color{black}(and in which sense)} individual extreme events to climate change \cite{Otto2017},} {\color{black} for defining science-based liability for their impacts. The  attribution of the 2010 RHW to climate change  has been heavily debated  \cite{Dole2011,Rahmstorf2011,Otto2012}.}  

\paragraph{This Letter} We aim at  advancing our understanding of 
heatwaves and cold spells in the Northern Hemisphere (NH) and of their response to  climate change. We treat the climate as a nonequilibrium system \cite{Lucarini2014b,Lucariniea2017,Ghil2020} and analyze it  
with Large Deviation Theory \cite{Varadhan1984,Touchette2009,Dembo1998} (LDT).  

LDT provides limit laws for the average of random variables, where stochasticity can also be due to deterministic chaos \cite{Kifer1990}. Let $A_n=\frac{1}{n}\sum_{i=1}^n X_i$, where the $X_i$'s are identically distributed, possibly correlated, random variables.   
$A_n$ obeys a Large Deviation Principle (LDP) if 
   $\lim_{n\to\infty}-\frac{1}{n}\ln p(A_n=a)=I(a)$
exists. 
$I(a)\ge 0$ is the rate function (RF), quantifying the exponential decay of probabilities with $n$ for all $a\neq a^*$, where  $I(a=a^*)=0$,  $\lim_{n\to \infty}p(A_n=a^*)=1$, and $a^*=\mathbb{E}[A_n]$. To verify the existence of a LDP for an observable having integrated autocorrelation time $\tau$ \cite{Billingsley2012} one can check whether 
\vspace{-2mm}
\begin{equation}
  I_n(a)=-\frac{1}{n/\tau}\ln p(A_n=a)
  \label{eq:ldest}
\vspace{-2mm}
\end{equation},
converges for {\color{black}larger and larger multiples of the averaging time $n=q\tau$, where usually $q\gg1$. In what follows both $n$ and $\tau$ are in units of days;  see supplemental material (SM)\footnote{The SM is accessible at \texttt{https://doi.org/10.6084/m9.figshare.14888151} and includes Refs. \cite{Osborn2014,r2016,brockwell2002}.}. 
{\color{black}LDT has been used to address some theoretical  aspects of geophysical fluid dynamics  \cite{Bouchet2012,Lucarini2014b,Bouchet_Rolland_Simonnet_2019:C} but has been otherwise not yet widely used in climate studies. Ref. \cite{Ragone2017} investigates heatwaves  by applying a genealogical algorithm to an intermediate complexity climate model; see also the recent follow-up work performed with a more complex climate model \cite{RagoneBouchet2021}. In \cite{Galfi2019} we construct rate functions for surface temperature (ST) data using a highly-idealized atmospheric  model.}  

{\color{black}Testing theories and methodologies across the model hierarchy is an effective research strategy in climate studies \cite{Held2005,Ghil2020}. In this work, we perform for the first time LDT-based analysis on the outputs of a state-of-the-art CMIP6 \cite{Eyring2016} Earth system model (ESM), namely the  MPI-ESM-LR model \cite{Giorgetta2013}. CMIP6 models have provided key inputs for the preparation of the latest (6th) report of the Intergovernmental Panel for Climate Change (IPCC).

We have two closely related main objectives:
\begin{enumerate}
    \item we will attempt to establish LDPs for the ESM output in order to define an LDT-based geographical climatology of heatwaves and cold spells and investigate their sensitivity to $CO_2$ concentration; 
    \item we will use the ESM to look into the 2010 RHW and the much less studied MD, and argue that these events, while extreme, are in some sense \textit{typical} and are part of the natural variability of the climate.
\end{enumerate}
A key element to address point 2. is that LDT captures the least unlikely of all the unlikely ways a large and persistent fluctuation can occur \cite{Hollander2000}. Let's use a  specific example that provides guidance for our analysis below. Hydrodynamic rogue waves can be explained using a LDP. As one takes sufficiently stringent height threshold criteria, the individual rogue waves become similar to each other, and their average converges to a special solution, associated with the so-called instanton \cite{Dematteis2019}. This formalism can be extended for treating events that have long duration and high intensity \cite{grafke2019numerical}, as in the  case  here.}

\paragraph{Rate Functions for the Surface Temperature}We first analyze large deviations of the ST in a 1000-year long pre-industrial control run (namely with fixed greenhouse gases concentration and land-use) of {\color{black}the MPI-ESM-LR model \cite{Giorgetta2013}. 
}
{\color{black}Building on the analysis of the climatic hot spots  \cite{Giorgi2006}, we consider the following regions in the NH: Northwest, Southwest, and Southeast America; the Mediterranean; North Europe; Northwest and Northeast Asia; the North Atlantic; and the North Pacific. }
We analyze summer and winter ST separately, in order 
to have seasonal data with nearly stationary statistics.  {\color{black}After removing the yearly cycle, 
we select for each year an extended summer - lasting $n_d=160$ days and beginning on May 5th - and an extended winter - lasting  $n_d=105$ days and starting on December 1st, see SM for details.}

We estimate RFs via Eq.~(\ref{eq:ldest}) for the spatially averaged ST in these regions and for the locales indicated in 
Fig.~1 of the SM. 
We compute  $I_n(a)$ for increasing averaging lengths $n=4\tau,8\tau,... m\tau=n_{max}\approx n_d$. 
The 
optimal averaging block length $n^*$ is such that $I_{n^*} \approx I_{n}$, $n>n^*$ \cite{Galfi2019}. We achieve convergence if $n^*\leq n_{max}$. 
The RFs for the regions are steeper than those of the corresponding locales and their convergence is faster thanks to spatial averaging \cite{Galfi2019}{\color{black}, as shown in Figs. 2-7 of the SM. It is very encouraging to see that LDT seems applicable at different levels of spatial granularity also in an ESM with realistic geography.} In all cases discussed below, the RFs are approximately quadratic. {\color{black}We will use this property at the end of the letter to present a preliminary example of the predictive power of our approach.} 

Figure~\ref{fig:surf} shows that for the summer RFs 
one finds convergence for the land areas and for the Mediterranean. 
The values of $n^*$ 
ranges 
between 1 and 2 months, which, encouragingly, corresponds with the time scale of actual high-impact heatwaves. 
The RFs are flatter for the North American and Eurasian regions, where a more continental climate with larger climate variability is observed, {\color{black}because a) the moderating effect of oceanic water masses is almost absent; and b) dry conditions can be more readily established and can lead to enhanced temperature fluctuations because of the reduced heat capacity of the soil.} 



The winter RFs - Fig.~\ref{fig:wirf} - are significantly flatter 
as an effect of the stronger atmospheric variability during winter; this is especially enhanced in continental regions, which feature large meridional temperature gradients \cite{Ghil2002}, so that the corresponding RFs are very flat. 
As the Mediterranean has a weak seasonality, the winter and summer RF for the ST are quite similar. 
The optimal averaging length also in this case 
ranges approximately between one and two months, which 
is compatible with the time scale of  cold spells in the real climate.

{\color{black}For both summer and winter ST, the RFs estimated for $n=n_{max}$ are very similar to those obtained by averaging over subsequent years, indicating that the LDP applies within a single season.  Finally, no LDP can be found over the ocean for either season. The basic difference in our ability to define RFs for ST over land and ocean agrees with the presence of long-term memory for the ocean ST  \cite{Fraedrich2003,Zhu2010}. Details on the convergence of the estimates of the RFs are shown in Figs. 8-9 of the SM.}
\begin{figure}[ht!]
\includegraphics[clip=true,trim=0.7cm 1.3cm .5cm 0.25cm,scale=.45]{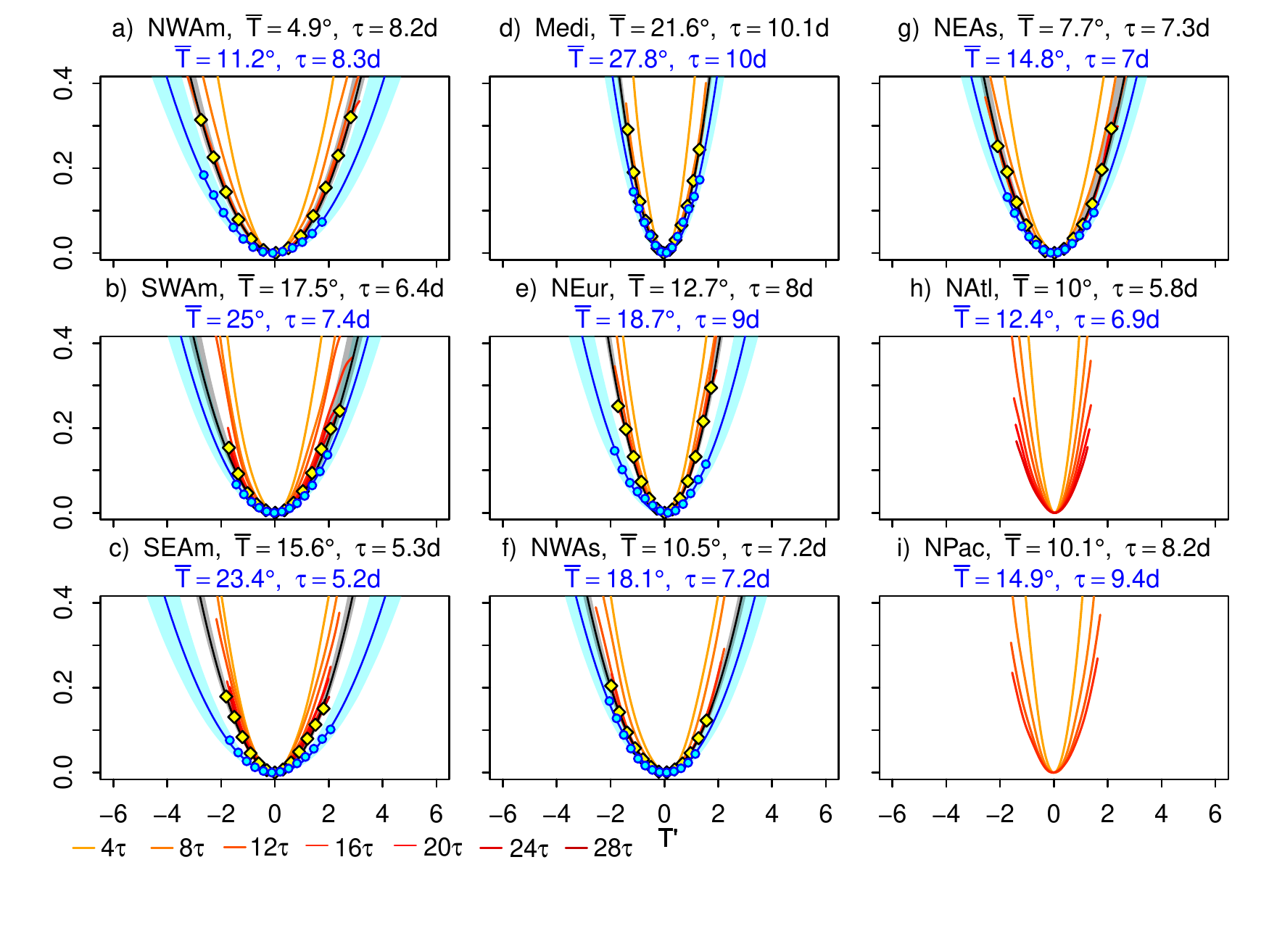}
\caption{\label{fig:surf} \small Summer RFs for the ST for increasing averaging windows (see legend) for a) Northwest, b) Southwest, and c) Southeast America, d) the Mediterranean, e) North Europe, f) Northwest and g) Northeast Asia, h) the North Atlantic, and i) the North Pacific. \textcolor{black}{The black (blue dashed) lines represents RFs obtained via \textcolor{black}{seasonal} averages for the pre-industrial (quadruple $\mathrm{CO_2}$) run, with thick (thin) lines showing empirical estimates (quadratic fits). On top the mean ST and $\tau$ for the control (black) and quadruple $\mathrm{CO_2}$ (blue) runs.} {\color{black}The 95\% confidence intervals of the rate functions - shaded areas - have been computed by bootstrapping the data year by year.}}
\end{figure}

\begin{figure}[ht!]
\includegraphics[clip=true,trim=0.7cm 1.2cm .5cm 0.25cm,scale=.45]{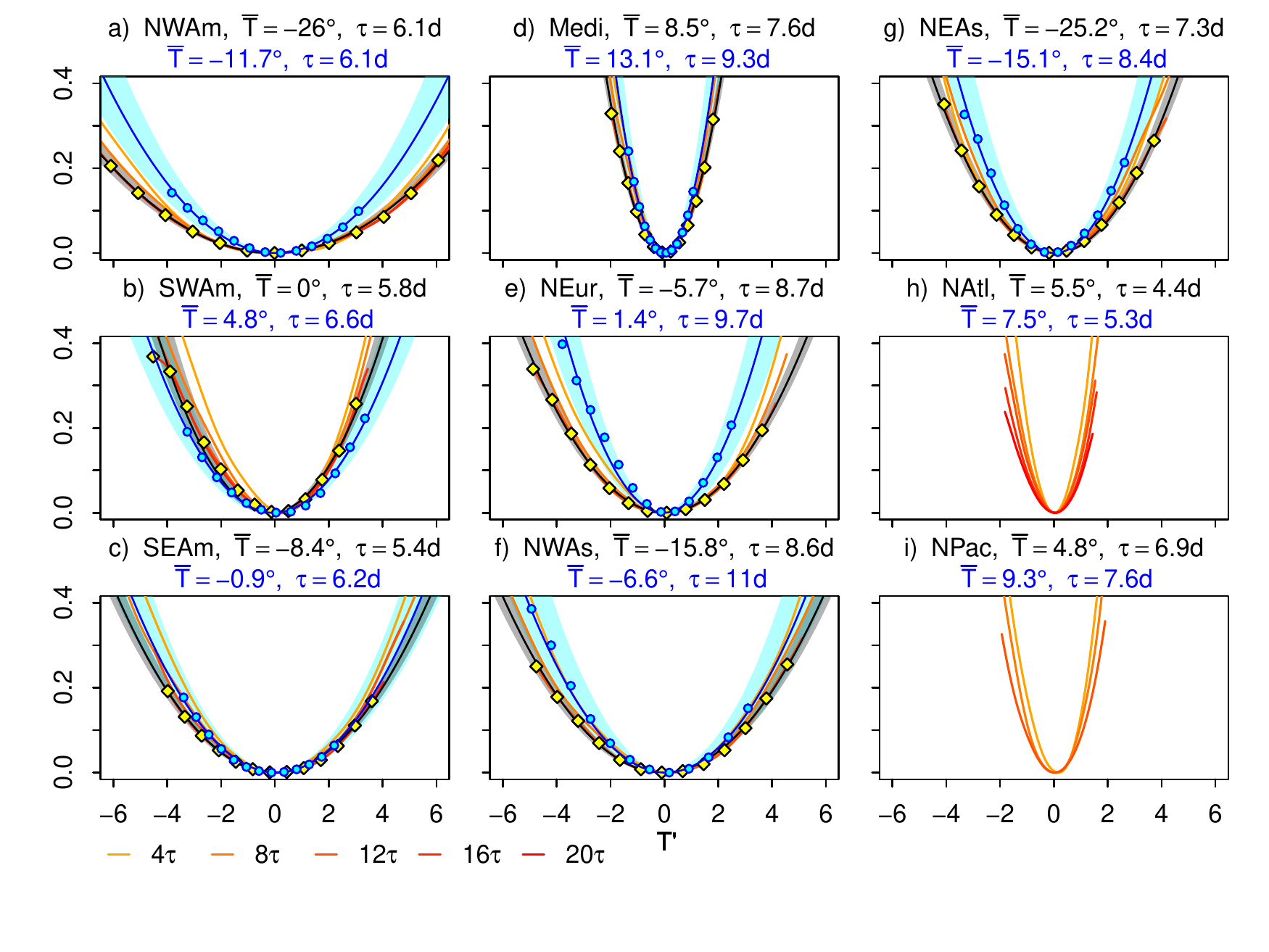}
\caption{\label{fig:wirf} Same as Fig.~\ref{fig:surf} but for winter RFs.}
\end{figure}
\paragraph{Heatwaves and Cold Spells in a Warmer Climate} 
{\color{black}We can infer, at least qualitatively, the impact of climate change on the statistics of heatwaves and cold spells by comparing the previous RFs 
with those computed by analyzing the ST fields of a 140-year long steady state simulation run with quadrupled CO$_2$ concentration.} 
This corresponds to a much warmer and more equitable climate, with a globally averaged ST higher by about 6 K and greatly reduced ST difference between low and high latitudes. 
The RFs for summer - see  Fig.~\ref{fig:surf} - are flatter {\color{black}in the Mediterranean and in all land regions}, thus indicating  an increased occurrence of heatwaves, also relative to much warmer average conditions. {\color{black}In some regions persistence is enhanced \cite{Kornhuber2019} as $\tau$ increases.} The increase in the probability of occurrence of heatwaves can be attributed to the drying of the soil, which 
activates a complex set of positive feedbacks \cite{Seneviratne2006}. Figure \ref{fig:wirf} shows that the winter RFs are everywhere steeper {\color{black}for the northern regions} in the warmer climate, 
 as the reduced ST difference between low and high latitudes leads to a weaker weather variability \cite{Screen2020} due to the reduced thermodynamic climate efficiency  \cite{Lucarini2014b}. {\color{black}Small changes are instead detected for the more southern regions.} Hence, considering the average ST increase, one expects fewer and less damaging cold spells in the future \cite{Smith2020}. {\color{black}See further details on RFs for the 4$\times$CO$_2$ simulation in the SM.}

\paragraph{Fingerprinting the 2010 RHW and MD}  
{\color{black} 

Our idea here is to use the viewpoint by \cite{Dematteis2019,grafke2019numerical} briefly presented above for interpreting the 2010 RHW and MD. The corresponding monthly (August and January 2010, respectively) mean fields of climatological anomalies  for} the ST and 500 hPa geopotential height (GPH) are reconstructed using the NCEP-NCAR reanalysis  \cite{Kanamitsu2002}; see the SM for additional ST maps from observations. The GPH arguably provides the most relevant information on atmospheric dynamics at synoptic and planetary scales \cite{Ghil2002,Ghil2020}. 
\begin{figure}[ht!]
\includegraphics[trim=0cm 0cm 13.2cm 0cm, clip=true, width=0.44\linewidth]{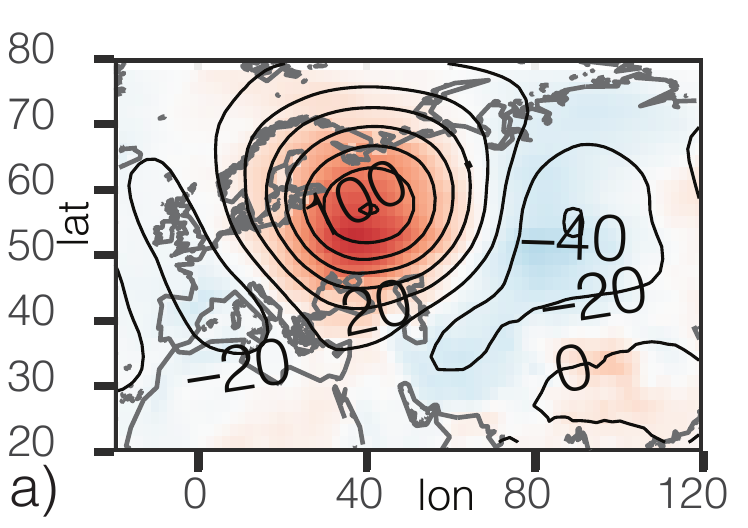}
\includegraphics[trim=11.cm 0.cm 0cm 0cm, clip=true, width=0.56\linewidth]{./figure3a.pdf}%
\caption{2010 RHW. a) Composite of ST anomaly fields from the model run with ST anomalies $\geq 4.5 K$ lasting 30 days in the locale indicated by the green dot. b) Mean August 2010 anomaly fields (NCEP/NCAR). The color map (isolines) indicates the ST (500 hPa GPH) anomalies. {\color{black}See in Fig. 11 of the SM the events contributing to the composite in a).} \label{fig:rhw}}
\end{figure}

\begin{figure}[ht!]
\includegraphics[trim=0cm 0cm 0cm 0cm,clip=true, width=0.98\linewidth]{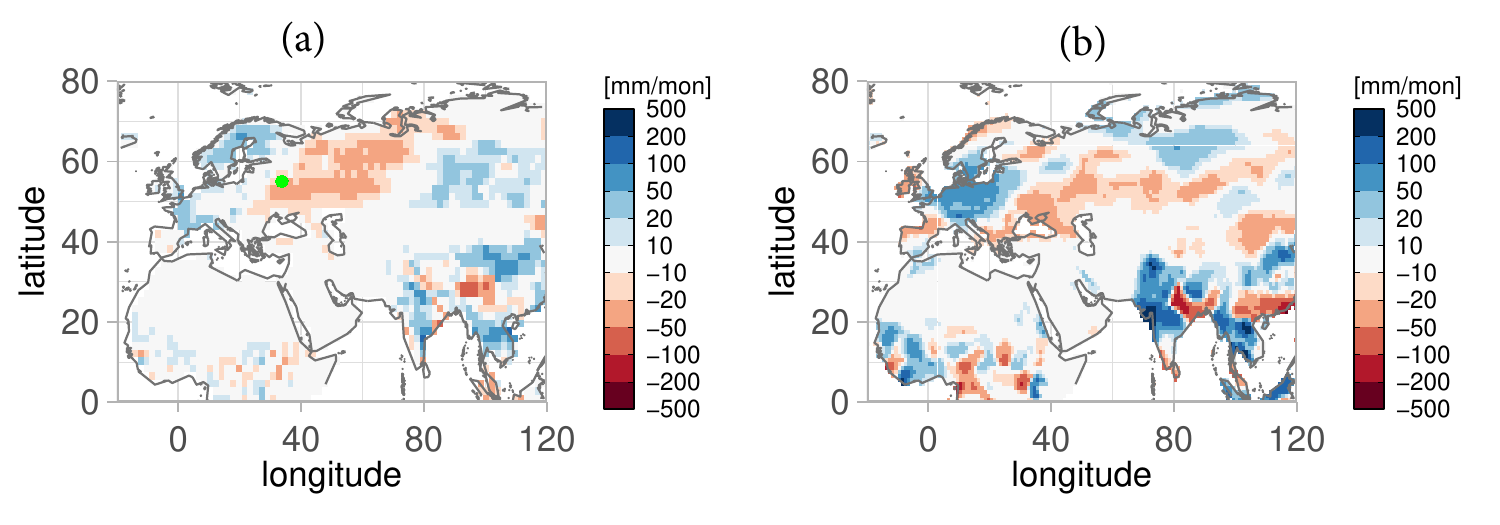}
\caption{2010 Russian Heatwave. a) Composite of precipitation anomaly fields constructed as in Fig. \ref{fig:rhw}. b) Mean August 2010 observed anomaly fields (CRU TS4 dataset). \label{fig:rhwprecip}}
\end{figure}

In order to investigate the RHW, we compute the mean anomaly fields for ST and GPH by averaging over the periods from the model run when we record ST anomalies  $\geq4.5$ K averaged over 30 days at the gridpoint located at {\color{black}$(55 \degree N, 33.75 \degree E)$ (west of  Moscow)}, which is in the core area of the 2010 event. Each of such periods (20 in total) corresponds to a heatwave. Size and duration of the fluctuations are chosen to match the observed one. 

The composite fields performed by averaging over the 20 heatwaves - Fig. \ref{fig:rhw} a) {\color{black} - portray our estimate of the solution associated with the instanton}. At large scales, these fields have a fair resemblance, both in shape and magnitude, with the observed anomalies for August 2010 
(Fig. \ref{fig:rhw} b). We see a similar pattern of positive ST and GPH anomalies over an extended circular region containing the selected grid point. This pattern is more symmetric and longitudinally less extended than its counterpart in the reanalysis data.  {\color{black}The spatial correlation between the fields is 0.54 (0.29) for ST (GPH), which is within the range of the value of spatial correlations computed between the individual heatwaves and their composite, which is $[0.47,0.87]$ ($[0.29,0.89]$) for the ST (GPH) field. {\color{black}Hence, the actual 2010 RHW can be seen as a fluctuation around the instantonic solution.} See Figs.~10-11 in the SM and comments therein.}  

{\color{black}The full GPH field} - {\color{black}see Fig. 10 in the SM -  shows that the composite captures a pattern that resembles the strong blocking high that is the well-known cause of the 2010 RHW and, indeed, a (rarely) recurrent local climatic feature \cite{Dole2011}.}   
{\color{black}{\color{black}Looking at the monthly cumulative precipitation (Fig.~\ref{fig:rhwprecip}), there is at least a qualitative agreement between the observed patterns of wet and dry anomalies  (panel b) and} the composite of the model data (panel a) in Europe, Western Siberia, parts of Africa, South, Southeast, and East Asia.} {\color{black}Despite a fair agreement over South Asia, 
the very intense wet spot in the Upper Indus basin is missing. This is hardly surprising given the well-know difficulty of  models in representing correctly the precipitation in that high-altitude locale \cite{Palazzi2015,Almazroui2020}. Given the complexity of the processes associated with precipitation, these results are encouraging.}



{\color{black}We proceed analogously for} the 2010 MD. We look in the model dataset for events featuring deviations of ST $\leq-10.5$ K averaged over 30 days at the gridpoint located at {\color{black}$(55 \degree$ N, 75 \degree$ E)$ (east of Omsk)}, which belongs to the core of the recorded event. {\color{black}We find 24 of such events}. 
The average fields of  ST and GPH recorded during the large deviations of the local ST (Fig.~\ref{fig:dzud} a) are in a {\color{black}fairly good} agreement 
with the  reanalysis data for January 2010 (Fig.~\ref{fig:dzud} b). {\color{black}The spatial correlation coefficient between the corresponding anomaly fields is 0.80 (0.64) for ST (GPH). These figures are, again within the range of the values of the spatial correlations computed between the model composite and the 24 realised heatwaves, which is $[0.62,0.90]$ and $[0.56,0.87]$ for the ST and GPH fields, respectively.  
 See also Figs. 12-13 in the SM and discussion therein.} 
 The spatial scale of  ST anomalies for the MD extends throughout Eurasia, with a large core region in Siberia and northern Mongolia. {\color{black}The full 500 hPA GPH field shown in Fig. 12 of the SM shows that also here  we can {\color{black}reconstruct the basic} mechanism behind the cold spell: a cut-off low in East Asia that leads to  advection of Arctic air into central Siberia.} 
\begin{figure}[ht!]
\includegraphics[trim=0.0cm 0.cm 13.7cm 0.0cm, clip=true, width=0.44\linewidth]{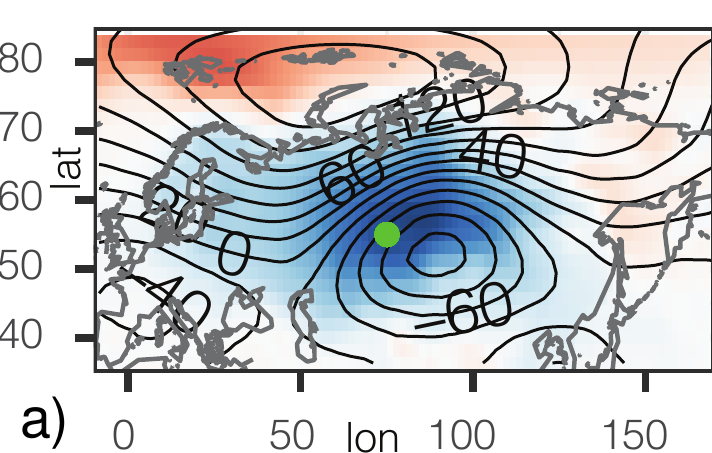}
\includegraphics[trim=11.5cm 0cm 0cm 0cm, clip=true, width=0.58\linewidth]{./figure5a.pdf}
\caption{Same as Fig. \ref{fig:rhw} but for the 2010 MD.  {\color{black}See in Fig. 13 of the SM  the  events contributing to the composite in a).} 
\label{fig:dzud}}
\end{figure}

\paragraph{Discussion {\color{black}and Outlook}} 
We have provided a new outlook on heatwaves and cold spells in the NH by applying LDT to the output of a state-of-the-art ESM. Extreme and persistent ST fluctuations during winter and summer over land and over the Mediterranean obey LDPs. The properties of the RFs quantify the regional and seasonal differences in the probability of occurrence of persistent extreme ST anomalies. We can also better appreciate the future risk due to heatwaves, as the increase in the average ST comes together with increased probability of such persistent positive fluctuations.

{\color{black}The obtained RFs are approximately parabolic in the range of practical interest.} Hence, the probability of observing an average anomaly $T_n$ of amplitude $a$ over a sufficiently long period of $n$ days is {\color{black} 
$\log\left(p(T_n=a)\right)\approx - n I_T(a)$, where $I_T(a)=a^2/(2\tau_T\sigma_T^2)$,}
where $\sigma_T^2$ is the daily variance and $\tau_T$ is 
in days.  
{\color{black}We can estimate the probability of occurrence of events of amplitude $a'$ and length $n'$ relative to that of less extreme events of amplitude $a$ and length $n$:}
\begin{equation}
p(T_{n'}=a')\approx 
p(T_n=a)\exp\left(\frac{na^2-n'a'^2}{2\tau_T\sigma_T^2}\right).\label{predic}
\end{equation}
{\color{black}We have attempted a first test of Eq. \ref{predic} on the RHW and MD locales,  
starting from the return time of $n=30$ days moderate events, i.e. 1$^o$C (-2$^o$C) for the RHW (MD). We can predict accurately the return times of  extreme events for $n=30$ days and $n=60$ days, see Fig. 14 in the SM. {\color{black}The formula applies \textit{a fortiori} for lower-resolution time series (e.g. weekly/monthly averages).} {\color{black}This idea, is promising for climate risk evaluation and deserves further study.}}

It is  intriguing that the anomalies of the ST and GPH fields during the  2010 RHW (MD) event look {\color{black}rather similar to those constructed by looking at the summer warm (winter cold) model ST anomalies selected conditionally to the presence of a large deviation of ST at the chosen locale. One constructs a large-scale pattern that is  involved in the occurrence of the persistent  event: a blocking over Russia (a deep low in East Asia) for the RHW (MD).} 
{\color{black}Using the conceptual framework proposed in \cite{Dematteis2019,grafke2019numerical},} we claim that these extreme events ({\color{black}our results are stronger for the MD}) 
are in fact  \textit{typical} - also at dynamical level, as part of the natural climate variability  - once we use the statistical lens defined by LDPs{\color{black}, which identifies the reference instantons.} Hence, they cannot be considered freak events or \textit{dragon kings} \cite{Sornette2012}.  
Similar conclusions were drawn for the RHW in \cite{Dole2011} through a more empirical yet informative approach. {\color{black}Ref. \cite{Otto2012} suggested the absence of an a-priori dichotomy between attributing the 2010 RHW to natural climate variability \cite{Dole2011} or to climate change \cite{Rahmstorf2011}.} {\color{black}Indeed, while its dynamics is part of the natural climate variability, its probability of occurrence is modulated by the changing climate.} 

{\color{black}The viewpoint proposed in this paper possibly contributes to understanding the low-frequency variability of the atmosphere \cite{Ghil2002,Ghil2020} and the role of stationary Rossby waves in causing  extreme events \cite{Lau2012,Kornhuber2019}, and is the starting point for further investigations of specific case studies. } 
{\color{black}In future work we will improve the quantitative evaluation of the agreement between observed extreme persistent observed and model-simulated events using tools like the Self Organizing Map \cite{kohonen2001}.
 One also needs to test the robustness of our findings by intercomparing different models, given the uncertainties on the skill of ESMs in representing the low-frequency variability of the atmosphere  \cite{Woolings2018} and its response to climate change \cite{Brown2020}.}  
\begin{acknowledgments}
The authors thank D. Faranda, G. Messori, F. Ragone, A. Speranza, and J. Wouters for many exchanges on extreme events and acknowledge the support by DFG TRR181 (grant no. 274762653). VL thanks B. Hoskins for having stimulated this investigation and 
acknowledges the support by the H2020 project TiPES (grant no. 820970). The authors have equally contributed to this study. 
\end{acknowledgments}


\providecommand{\noopsort}[1]{}\providecommand{\singleletter}[1]{#1}%

\end{document}